The emergence of the X-ray luminosity/cluster richness relation for radio galaxies

David Garofalo[1] & Chandra B. Singh[2]

1. Department of Physics, Kennesaw State University, Marietta GA 30060, USA
2. South-Western Institute for Astronomy Research, Yunnan University, University Town, Chenggong, Kunming 650500, People's Republic of China

*Abstract*

The idea that mergers are more likely in dense groups or clusters coupled with the assumption that such events lead to cold gas flows onto black holes, suggests a direct relationship between the radiative efficiency of an active galactic nucleus and environmental richness. Observations, however, increasingly challenge this and other basic expectations. Mounting evidence, for example, shows an inverse trend between near-Eddington accreting objects and environmental richness. Broken down by radio galaxy subgroup, recent work has explored connections between low excitation radio galaxies with Fanaroff-Riley II jet morphology (FRII LERGs) and other radio galaxies. We make contact with that work by adding a discussion of the recently discovered FR0 radio galaxies and show how to fit them in a picture in which FRII LERGs are not initial nor final phases in the lifetime of a radio galaxy, but de facto transition states. We describe how to understand the observed X-ray luminosity/cluster richness relation as a fundamental correlation on the nature of the jet-disk connection.

1. Introduction

Around 1% of galaxies (having black holes in the range of $10^6 – 10^9$ solar masses) exhibit activity in their centers or nuclei and are termed active galactic nuclei (AGNs). About 10% of AGNs sometimes give rise to the formation of relativistic jets emitting non-thermal radiation. These sources are called radio-loud AGNs (RLAGNs) or jetted AGNs (Padovani 2016). RLAGNs are classified as blazars (jet pointed towards the observer) and radio galaxies (RGs – with jets pointed close to the sky plane). Fanaroff & Riley (1974) further classified RGs as FR I (edge-darkened RGs) and FR II (edge-brightened RGs). RGs are also classified by their excitation class: high-excitation (HERGs) and low-excitation (LERGs). HERGs are found to be in radiative or quasar mode (most of the AGN energy output in accretion) with radiatively efficient disk luminosity greater than 0.01 the Eddington luminosity ($L_{Edd}$) while LERGs operate in jet mode (most of the AGN energy output in the jet) with radiatively inefficient disk luminosity less than $0.01 L_{Edd}$. LERGs tend to have very massive black holes, FR I or II radio morphology, weak (or absent) narrow, low-ionization emissions lines while HERGs have a massive black hole, overwhelmingly FR II morphology, and strong high-ionization narrow lines (Heckman & Best 2014 and references therein).

Recent observations have led to the emergence of an anomalous correlation between the X-ray luminosity of different classes of radio galaxies and their cluster richness (Macconi et al 2020). The low numbers and recent origin of this correlation imply that a deeper study is required before it is widely accepted. Despite the tentative nature of the correlation, we argue that it will emerge as fundamental to the behavior not merely of radio galaxies, but of the AGN phenomenon more generally. Specifically, we will show that AGN capable of coupling strong jets with bright disks are short lived in a way that is black hole mass dependent, and because larger black holes live in different environments compared to those with smaller masses, this translates into environmental dependence. Because of the time evolution in the paradigm, we show that from an observational perspective, brighter accreting black holes will tend to be found in relatively less rich environments. These ideas emerge from a paradigm for black hole accretion and jet formation that introduces counterrotation between accretion disk and black hole as the cornerstone of the jet-disk connection (Garofalo, Evans & Sambruna 2010). In addition to explaining the nature of the X-ray luminosity/cluster richness relation and the location on the relation of the different subclasses of radio galaxies, we will be able to make predictions. In particular, we will show how to incorporate the recent class of FR0 radio galaxies onto this correlation as intermediate or transition objects by appealing to an idea from a recent study on the environmental dependence of FR0 radio galaxies compared to FRI radio galaxies (Capetti, Massaro, Baldi, 2020).

Macconi et al (2020) adopt the definition of cluster richness (CR) used by Gendre et al (2013) who followed the method of Wing & Blanton (2011) based on counting galaxies within a 1 Mpc radius disk. This differs from the method for calculating CR from Capetti et al (2020) who use a radius of 2 Mpc for "cosmological neighbors" (Massaro et al 2019). In this work we will make predictions about the CR values of FR0 radio galaxies based on the method adopted by Gendre et al (2013) but note that if one adopts the methods of Massaro et al (2019), all CR values should shift to higher values due to the larger area on the sky for inclusion in the CR value. FR0 radio galaxies are a recently discovered class of radio galaxies with compact jets that dominate the radio galaxy population at low redshift (Capetti, Massaro, Baldi 2017a; Capetti, Massaro, Baldi 2017b; Baldi, Capetti, Massaro, 2018; Baldi, Capetti, Giovannini, 2015; Baldi, Capetti, Giovannini 2019; Baldi & Capetti 2009; Baldi & Capetti 2010). While the sources of Macconi et al. are from the 3CR catalog whose radio flux limit is on the order of Jy, FR0s together with FRICat and FRIICat sources (Baldi et al. 2018, Capetti et al. 2017a,b) are selected at lower radio fluxes (on the order of mJy) and at a different redshift: 3CR sources are all at $z < 0.3$, FRICat and FRIICat are at $z<0.15$, while FR0s are at $z < 0.05$.

In Section 2a we describe the expectations given our current understanding of the triggering of jetted AGNs and in Section 2b we discuss actual observations. In Section 2c we apply our model to explain the observations and predict the distribution of FR0 radio galaxies and in Section 3 we conclude.

2. *Discussion*

*a. Expectations*

The basic picture for the triggering of powerful jetted AGN is a gas rich merger leading to cold gas accretion onto massive black holes in galactic centers (e.g. Barnes & Hernquist 1991). This cold gas would accrete onto black holes at high rates and generate strong feedback (e.g. Silk & Rees 1998). This simple picture predicts that the most massive cold mode accreting black holes will produce the most powerful feedback. However, this picture fails the observations in many respects. The most powerful active galaxies are known as FRII quasars, a combined term that refers to the type of jet and accretion. As mounting observations indicate, FRII quasars do not have the most massive black holes compared to active galaxies whose average feedback is weaker. This basic failure has led to the recent idea that mergers in hot clusters may trigger streams of cold gas onto black holes (Hardcastle 2018). Unfortunately, the evidence for FRII quasars in rich clusters is weak. In fact, not only do FRII quasars appear not to share the most massive black holes, they are also found in increasingly isolated environments (Magliocchetti et al. 2018; Macconi et al. 2020).

FRI radio galaxies possess the most massive black holes, are overwhelmingly found in low excitation states and at low redshift, but they are outnumbered 5 to 1 by FR0 radio galaxies at z < 0.05. These objects need to be reconciled with the Soltan argument, which prescribes the accretion efficiency at low redshift to be high and that black holes at low redshift must be spinning rapidly. The observation of powerful jets at high redshift with the high accretion efficiency estimated at lower redshift produces tension with models of AGN (the Meier paradox) because such models tend to associate both strong jets and high accretion efficiency with high black hole spin. Resolving this issue is at the foundation of black hole astrophysics. What is lacking is a picture that makes sense of the most powerful active galaxies in terms of their black hole masses, black hole spins, redshift, environment, and excitation class, as well as a connection to the rest of the active galaxy phenomenon. We will illustrate a framework that accomplishes this for the entire radio galaxy population that also allows for an understanding of the transitional nature of the recently discovered FR0 radio galaxies.

*b. Observations*

Macconi et al. (2020) have explored the X-ray luminosity and richness factors of FRI LERGs, FRII LERGS and FRII HERGs. The data are plotted in Figure 1. We assume that at least most of the X-ray luminosity is generated by processes that are directly related to accretion and therefore that X-ray luminosity scales with disk efficiency and serves as a proxy for accretion efficiency (this is confirmed by the accretion rate as estimated with [OIII] emission line, see Macconi et al. 2020). However, some caution is necessary here since part of the non-thermal jet spectrum appears in X-rays in FRI LERGs (Balmaverde et al 2006) and in FR0 LERGs. Therefore, the accretion luminosity must be less than the measured Lx in the 2-10 keV range. FRII HERGs are found to occupy the highest values of Eddington scaled X-ray luminosity, yet their richness factor is low, with the majority of objects leftward of the boundary line between poor and rich environments. These are shown in blue. As mentioned in Section 1a, this does not obey the expectation that the largest merger products should produce the largest inflow of cold gas onto largest black holes which reside in rich cluster environments (Hardcastle 2018). Subsamples of FRII HERGs, however, have larger CO measured black holes compared to FRI LERGs and appear to be merger triggered (Ocana-Flaquer et al 2010; Smolcic & Riechers 2011). Although the numbers in Figure 1 are small and the difference in black hole masses is statistically insignificant, FRII HERGs are known *not* to be characterized by the most massive black holes which tends to fit alongside the fact that they are observed in poorer environments. As we will argue in Section 1c, we can understand FRII HERGs as merger-triggered (Chiaberge et al 2015), cold mode accreting, initial states in the radio galaxy phenomenon while other radio galaxies constitute latter stages, and therefore it makes sense that on average they should have the lowest black hole masses. In the FRIICAT sample (Capetti et al. 2017), the FRII black hole masses are statistically and significantly less massive than FRII LERGs. In fact, in the FRIICAT sample the mean values for the logarithm of the mass ratios are 8.13 for FRII HERGs and 8.37 for FRII LERGs. A Kolmogorov-Smirnov (K-S) test confirms that the two samples are drawn from two different populations with a probability larger than 99.9%. In FRIICAT the black hole masses are calculated in a different way compared to Macconi et al., which prevents any direct comparison.

Compared to FRII HERGs, FRII LERGs occupy both a lower Log ($L_x/L_{Edd}$) range of values, and are shifted slightly toward the richer environments. They are plotted in green. Lastly, in red, are FRI LERGs which appear at the lowest Log ($L_x/L_{Edd}$) range and highest CR values. As we will show in Section 1c, FRII LERGs are a possible future state of an FRII HERG, but that probability increases with cluster richness. As a result of nature of the time evolution of FRII HERGs, as we will see, with respect to FRII HERGs, FRII LERGs should not only shift downward, but also to the right on the diagram of Figure 1, where they in fact are observed.

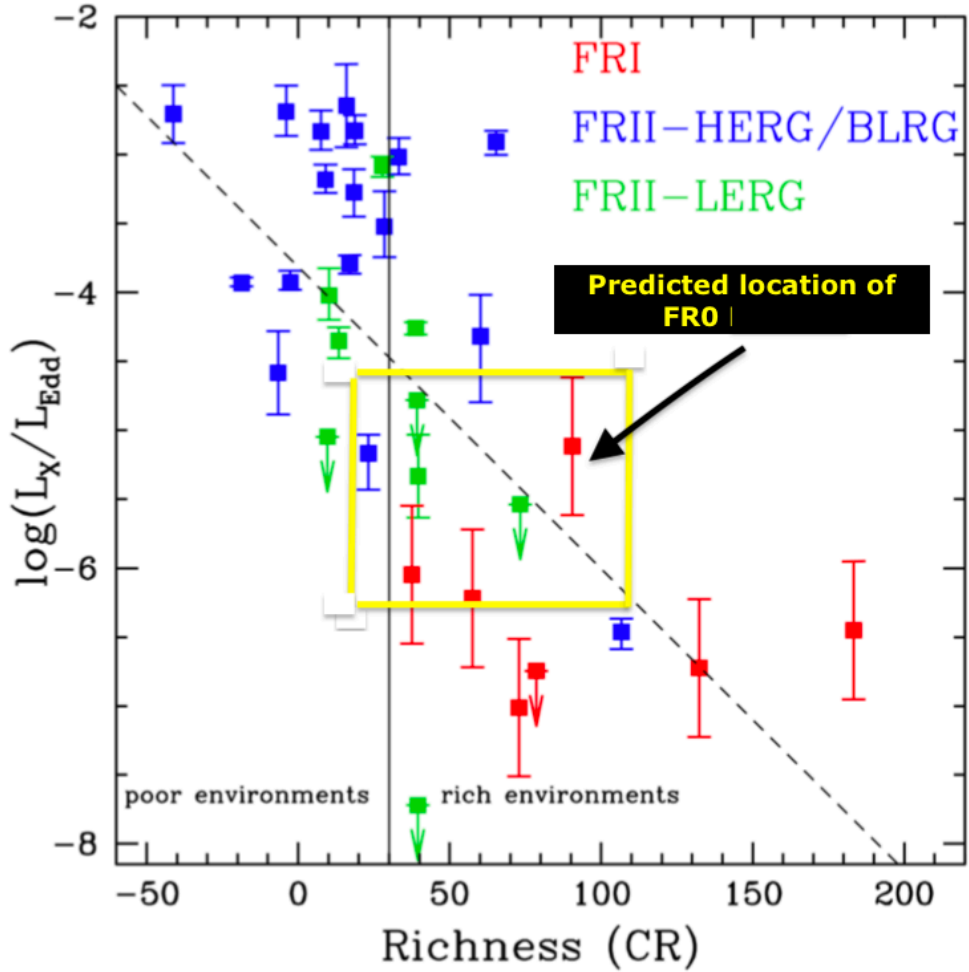

Figure 1: Observed X-ray-scaled Eddington luminosity versus cluster richness (CR) for FRI LERGs, FRII HERGs/LERGs from Macconi et al. (2020) including the location of FR0 radio galaxies expected from the model.

As we will describe, the theoretical picture that explains Figure 1 will also predict a distribution of FR0 radio galaxies on this plot such that they are sandwiched between FRII LERGs and FRI LERGs. We will argue that they, like FRII LERGs, are in fact transition objects, but come into play at later times as progenitors to the FRI LERGs.

### c. Explanations

In Figures 2-4 we show the basic time evolution of cold mode accreting black holes in retrograde configuration which enhances the jet mechanism and thus leads to an FRII morphology. It is important to point out that retrograde accretion constitutes much less than ½ of all accretion states (see Garofalo, Christian & Jones 2019 and Garofalo, North, Belga & Waddell 2020 for details). Black hole mass increases from Figure 2 to Figure 4 which also makes jet power increase. More massive black holes live in more massive dark matter halos which associates to richer environments. Hence, environment richness increases from Figure 2 to Figure 4. Because of the greater jet feedback, radio galaxies described in Figure 4 change their accretion state rapidly while this occurs less quickly or not at all in less rich environments. The greater jet feedback results in the model from the combined effort of the Blandford-Znajek and Blandford-Payne mechanisms (Blandford & Znajek 1977; Blandford & Payne 1982). The former mechanism depends on the spin of the black hole – which does not by itself prefer retrograde over prograde configurations – the black hole mass, and the magnetic field threading the black hole. Because accretion disks are diffusive, retrograde configurations bring stronger magnetic fields to the black hole and this selects the retrograde configuration as the one with more powerful Blandford-Znajek effect (Garofalo 2009). The Blandford-Payne mechanism, on the other hand, depends on mass loading onto magnetic fields lines threading the disk, which hinges on the bend of the poloidal field toward the disk. This bend increases in the retrograde configuration due to the greater magnetic flux bundle on the black hole which serves to bend the disk-threading magnetic field via magnetic pressure. While the former mechanism prescribes strong jets in retrograde configurations, weak jets for low spins, and intermediate jet power for prograde spins, the latter mechanism is more effective as the disk is closer in its spin to higher retrograde values. Overall, the jet power dominates in the retrograde configuration. The evolution of the disk depends on ISM heating by the jets which occurs more effectively for the most massive black holes (see Garofalo, Evans & Sambruna 2010 for details).

Figure 2 shows an FRII HERG transitioning to an FRI HERG. The timescale for that process reduces to evaluating the amount of mass needed to spin the black hole up into the prograde regime to a value of 0.2. This means the black hole accretes from its original mass $m_0$ in solar masses, to a mass of $1.3m_0$ (Kim et al 2016). The model assumes the initial accretion occurs at the Eddington limit, which allows us to evaluate timescales for transition through different black hole spin values as given by

$$\int_0^T \frac{dm}{dt} dt = m \qquad (1)$$

which in this case looks as follows.

$$\int_0^T \left(\frac{dm}{dt}\right) dt = 0.3 m_0 \qquad (2)$$

where $dm/dt = 2.2 \times 10^{-8}\ m_0$. For simplicity, we assume $dm/dt$ is constant. This means the black hole accretes at the Eddington limit only at t=0 but drops slightly but continuously below it as the mass of the black hole increases thereafter. This allows us to extract the accretion rate to obtain

$$(dm/dt)\ T_1 = 0.3 m_0 \qquad (3)$$

which results in a time

$$T_1 = 0.3 m_0 / (dm/dt) = 0.3 m_0 / 2.2 \times 10^{-8}\ m_0 = 1.35 \times 10^7\ \text{years.} \qquad (4)$$

To reach the high prograde spin of 0.9 from the original high spin retrograde configuration, instead, requires accreting to a mass of about twice the original black hole mass. The timescale for that process is therefore given by

$$(dm/dt)\, T_2 = 2m_0 \qquad (5)$$

$$T_2 = 2m_0/(dm/dt) = 2m_0/2.2 \times 10^{-8}\, m_0 = 9 \times 10^7 \text{ years.} \qquad (6)$$

If accretion rates drop to below 0.01 of the Eddington accretion rate, we have advection dominated accretion (ADAF) and timescales that, compared to the Eddington timescale, increase by at least a factor of 100. Based on these ideas, we can estimate timescales for the different phases in Figures 2-4 which we report in Table 1. To obtain these timescales one applies equation (1) with the appropriate final mass as given in Kim et al (2016) which amounts to some fraction of $m_0$.

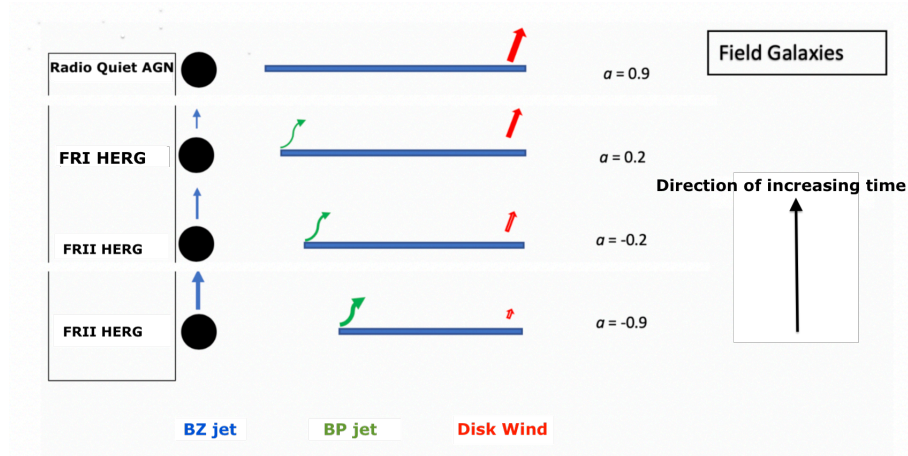

Figure 2: Time evolution of originally retrograde accreting black hole in cold mode (lowest panel). BZ represents the Blandford-Znajek jet, BP the Blandford-Payne jet, and the red arrow represents the radiatively driven disk wind which gets its power from the inner disk region through a reprocessing of that energy further out in the disk. Time increases upwards as indicated by the arrow. Negative spin represents retrograde configurations while positive spin represents prograde ones. Environment is labeled at the top right while radio morphology and excitation class is on the far left. Since smaller dark matter halos occupy isolated environments, their black holes are smaller and their feedback weaker. Hence, their accretion mode is weakly affected and does not change appreciably from the cold mode.

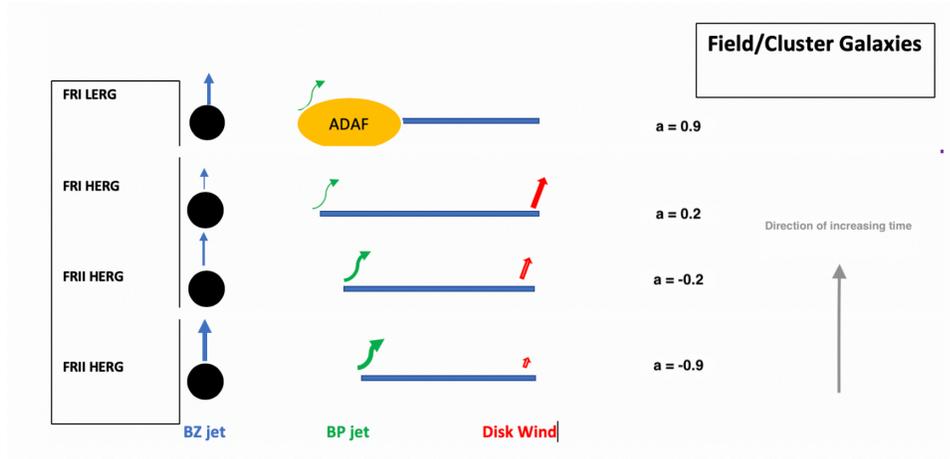

Figure 3: Same as Figure 2 but for systems with larger average black hole masses whose jet feedback is more effective in altering the mode of accretion. At late times these systems accrete hot gas in advective dominated flows.

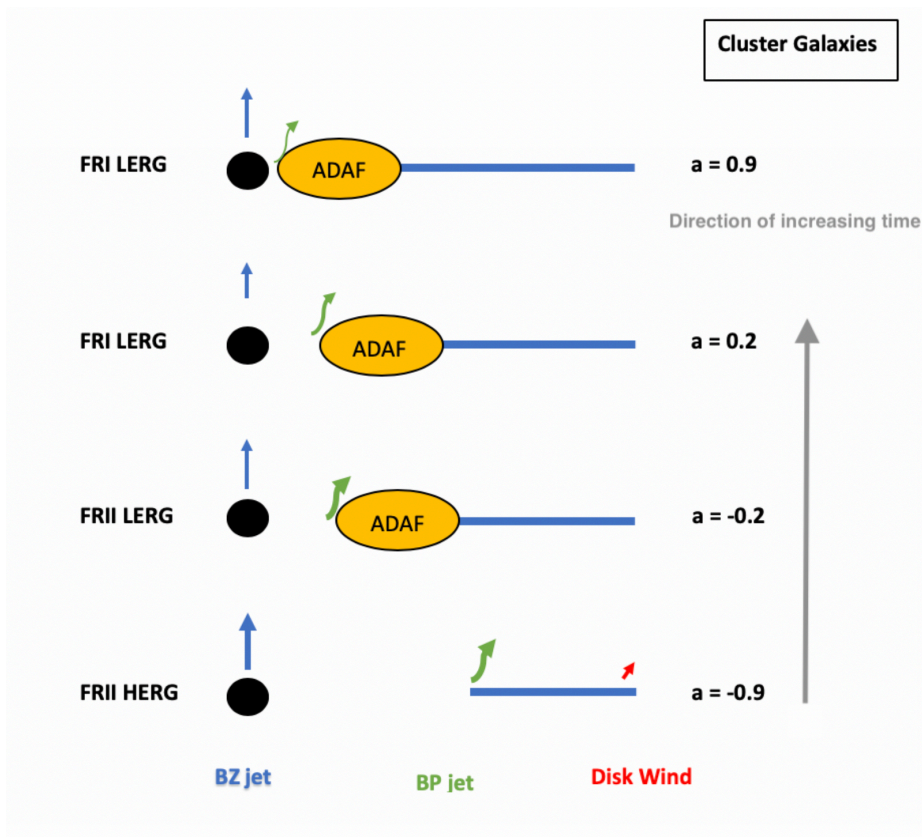

Figure 4: Same as Figures 2 and 3 but for the most massive black holes on average which tend to dominate in richer environments. Because these black holes are the most massive, they tend to produce the most powerful and effective jet feedback, which rapidly alters the state of accretion.

| Family | Avg. Lifetime (rich/poor) |
|---|---|
| FRII HERG | $4 \times 10^6$ yrs./ $8 \times 10^6$ yrs. |
| FRII LERG | $4 \times 10^8$ yrs./ smaller |
| FR0 HERG | smaller/ $10^6$ yrs. |
| FR0 LERG | $10^8$ yrs./smaller |
| FRI HERG | smaller/a few $10^6$ yrs. |
| FRI LERG | $10^{10}$ yrs./greater than zero |

Table 1: Timescales for lifetimes of different radio galaxy subgroups that make sense of the location of FR0s in Figure 1. The times depend on jet power and accretion rate. For a HERG in a poor environment, the accretion rate is within the range $0.01(dm/dt)_{Edd} < dm/dt < (dm/dt)_{Edd}$ but that tends to be less than in a rich environment. And for a LERG in a poor environment, the state will tend to last less than in a rich environment because a greater amount of fuel tends to be accreted in a HERG state prior to the LERG state, which then has less fuel to feed from and a shorter lifetime.

- *FRII HERGs*

Why do FRII HERGs have higher radiative efficiency compared to FRI LERGs? Is there more gas available in the FRIIs? If so, then why is the CR factor in FRII HERGs less than in FRI LERGs? And what is the reason for the FRII morphology? Could it be due to a less dense environment? That would seem to fit with the lower CR factor. If mergers are greater in regions of higher CR factor and they lead to cool gas flows onto black holes, we expect higher radiative efficiency to correlate with higher CR factor. But Figure 1 appears to suggest that black holes experiencing abundant cold gas flow live in environments that experience fewer mergers. How then can we make sense of the position of FRII HERGs in Figure 1? Figures 2 and 4 allow us to understand this. In the less rich environment (Figure 2), the merger triggers the FRII HERG in a black hole system whose jet feedback is relatively weaker and incapable of changing its accretion state, which persists in cold mode. These are the longest lasting FRII HERGs. In rich environments (Fig.4), by contrast, the greater jet feedback will change the accretion state rapidly, which makes FRII HERGs in rich clusters the shortest living. Everything else being equal, the probability of detecting an FRII HERG in a poor environment is therefore twice that of detecting it in a rich environment. However, merger rates are environment-dependent so the probability of finding an FRII HERG in a poor environment would be less than twice that of finding one in a rich environment. Despite the difference in likelihood of finding an FRII HERG in different environments, Figures 2-4 indicate that the most luminous FRII HERGs with the more massive black holes are predicted to live in richer environments. But in such environments, they last less on average. For the objects of Figure 1, cluster richness versus black hole mass show no trend so this is a prediction that should be checked as the numbers increase.

- *FRII LERGs*

Based on nomenclature, FRII LERGs differ from FRI LERGs in their jet morphology. But what is the cause for the generally more powerful jets in the former? Considering black hole mass, on average FRI LERGs have the largest values so that appears ruled out. Black hole spin detection is only in its infancy so we cannot constrain that observationally. Perhaps accretion rates are greater in FRII LERGs and larger accretion rates drag stronger magnetic fields to the black hole. Like for FRII HERGs, tension appears here in that higher accretion rates are expected after mergers, suggesting that FRII LERGs would have to be found in richer environments compared to FRI LERGs. Just as for the FRII HERGs, the opposite is true, as seen in Figure 1. This suggests that FRII LERGs are not the direct product of a merger but rather a phase in the lifetime of a black hole that was originally triggered in a merger.

Our explanation for the position of FRII LERGs in Figure 1 emerges from Figures 3 and 4. Only for systems whose jet feedback is powerful enough during the FRII HERG state will feedback rapidly alter the accretion mode into an ADAF. If the system is still spinning in retrograde configuration, the jet is of FRII morphology but is weaker than in its FRII HERG state due to the black hole spinning down. Since FRI LERGs are in the future of FRII LERGs, the black holes of FRI LERGs have accreted more than 20% of the mass they had when they were born as FRII HERGs. And the jet morphology has switched because accretion has transitioned into the prograde regime which the model prescribes to be less effective due to less magnetic flux and weaker Blandford-Payne jets. Our picture also prescribes lower redshift values for FRI LERGs compared to FRII LERGs, which is compatible with observations (Laing et al., 1983; Spinrad et al. 1985; Wall et al. 1985; Morganti et al. 1993; Burgess et al. 2006). In more isolated environments (i.e. Figures 2 and 3), FRII LERGs either do not form at all, or they form with comparably less total available cold gas compared to rich clusters where the biggest mergers occur. The less gas available, the less of the FRII LERG to FRI LERG evolution is experienced so one expects more radio galaxies in isolated environments to struggle to reach the FRI LERG state. If they do reach the FRI LERG state, however, they would not live long in that state and rapidly become dead quasars as fuel runs out. In the richest environments, on the other hand, the gas is plentiful, and FRI LERG states are the longest lived of any radio galaxy. The timescale for FRII LERGs in Table 1 comes from our understanding of Eddington-limited accretion applied to a black hole with intermediate spin in retrograde configuration. Since the Eddington-limited timescale to spin such a black hole down to zero is about $4 \times 10^6$ years, and that an ADAF state accretes at a rate

$$dm/dt < 0.01(dm/dt)_{Edd},$$

we can obtain a timescale. Given that the ADAFs in these FRII LERGs have recently undergone the transition from a radiatively efficient disk, it is reasonable to use 1% of the Eddington accretion rate which produces a timescale for the FRII LERG of $4 \times 10^8$ years. Since ADAF accretion states result from effective jet feedback, isolated environments will enter the ADAF state later, if at all (i.e. Figure 3). Hence, the average timescale for FRII LERGs in isolated environments must be lower than the $4 \times 10^8$ years (note that no FRII LERGs appear in Figures 2 and 3, suggesting they are not dominant in non-rich environments). The idea of a recent transition to an ADAF or LERG state in FRII LERGs allows us to appreciate the intermediate radiative efficiency of FRII LERGs compared to higher radiative efficiency in FRII HERGs and lower radiative efficiency in FRI LERGs. The latter, in fact, come into being at later times as accretion rates drop further. If the fuel is insufficient (poor environments), however, FRII LERGs will not evolve into FRI LERGs. In other words, while FRI LERGs tend to be the dominant phase for radio galaxies in richer environments, and therefore will be observed in richer environments, not so in isolated ones where they struggle to form. Because of their tendency to be absent in poor environments, therefore, FRI LERGs will be observed in rich environments while FRII LERGs will be found in a range of environments that includes rich and poor. In short, the time evolution prescribed in Figures 3 and 4 explains the intermediate, transitional nature, of FRII LERGs in Figure 1.

Finally, we compare the predictions of this model with the idea that ADAF disks could be induced by the hot cluster gas independently of jet feedback (Hardcastle 2018). In this picture, one might expect the more powerful FRII LERGs to form toward the cluster center with larger black holes, while FRI LERGs might form further away, and with smaller black holes; and, with no average difference in redshift between the two groups. Instead, FRII LERGs generally have higher average redshifts and lower average black hole masses compared to FRI LERGs. As end-of-the-line offspring, our model allows us to appreciate why FRI LERGs emerge at lower redshifts, with the largest black holes, and in the richest environments.

*FR0 LERGs*

FR0 LERGs are represented in our Figures and have recently been found to outnumber FRI LERGs at low redshift (Baldi & Capetti 2009, 2010; Baldi et al 2018). Although they are not directly present in either Figure 3 or 4, we can infer their location by their jet morphology and excitation state. As accretion spins FRII LERGs down, jet power decreases, disappears at zero spin, and re-emerges as the spin $a$ satisfies $a > 0.1$ in the prograde direction. This weak prograde jet has been modeled as an FR0 (Garofalo & Singh 2019) which lasts until the spin is $a \sim 0.2$, after which it becomes an FRI. We show this in Figure 5 which amounts to a more detailed description of the transitions in the richest environments. While such evolution tends to dominate in richer environments, there are similar systems in less rich environments that become ADAFs in their accretion state a little later in time, at or near the zero spin transition. For simplicity, Figure 3 captures the idea that ADAF accretion states come into play later by introducing an ADAF accretion state above spin value of $a = 0.2$ but there should be a continuous range of systems that produce ADAF accretion after the retrograde regime and these will tend to occur in less rich environments. Consider accreting black holes that are surrounded by ADAF disks during the period the spin satisfies $0.1 < a < 0.2$. These are FR0s in their jet morphology and will become FRIs if enough fuel exists. However, FR0s in which less fuel exists and where accretion rates are lower, are the ones that will last longer, and will therefore be more likely to be observed. As pointed out by Capetti, Massaro & Baldi (2020), FR0s in rich clusters last shorter times compared to FR0s in groups because the latter tend to linger in the $0.1 < a < 0.2$ range for longer which explains the tendency to observe FR0 LERGs closer to the boundary between rich and poor environments compared to FRI LERGs as seen in the yellow box of Figure 1. We plot the FR0s of Massaro et al (2019) and show them in Figure 6 in black. Some caveats need to be discussed, however. First, these FR0s are found in a fainter radio selected catalog compared to the 3C catalog which forbids a direct comparison for the accretion rates. The FRII-LERG and FRII-HERG accretion rates obtained with [OIII] line luminosity in the FRII-CAT sample (Capetti et al 2017) are about one order of magnitude lower with respect to the 3C sample. Hence, the lower FR0 accretion rates in Figure 6 has a systematic offset but one that we cannot quantify. Second, their cluster richness classification is obtained using a different method compared to Gendre et al (2013) as Massaro et al (2019) consider as companion sources only galaxies with spectroscopic redshift measurement, while Gendre et al (2013) include photometric redshift measurements, a systematic difference that likely results in a shift of the plotted FR0s toward smaller CR values. Evaluating these systematic differences is beyond our expertise so we leave it to the observers to further explore the degree to which the black points on Figure 6 migrate away or toward the yellow boxed region.

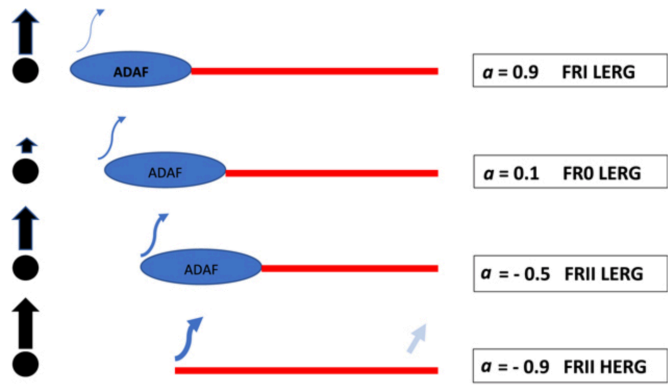

Figure 5: Detailed time evolution of FRII HERGs in richer environments. From Garofalo & Singh (2019).

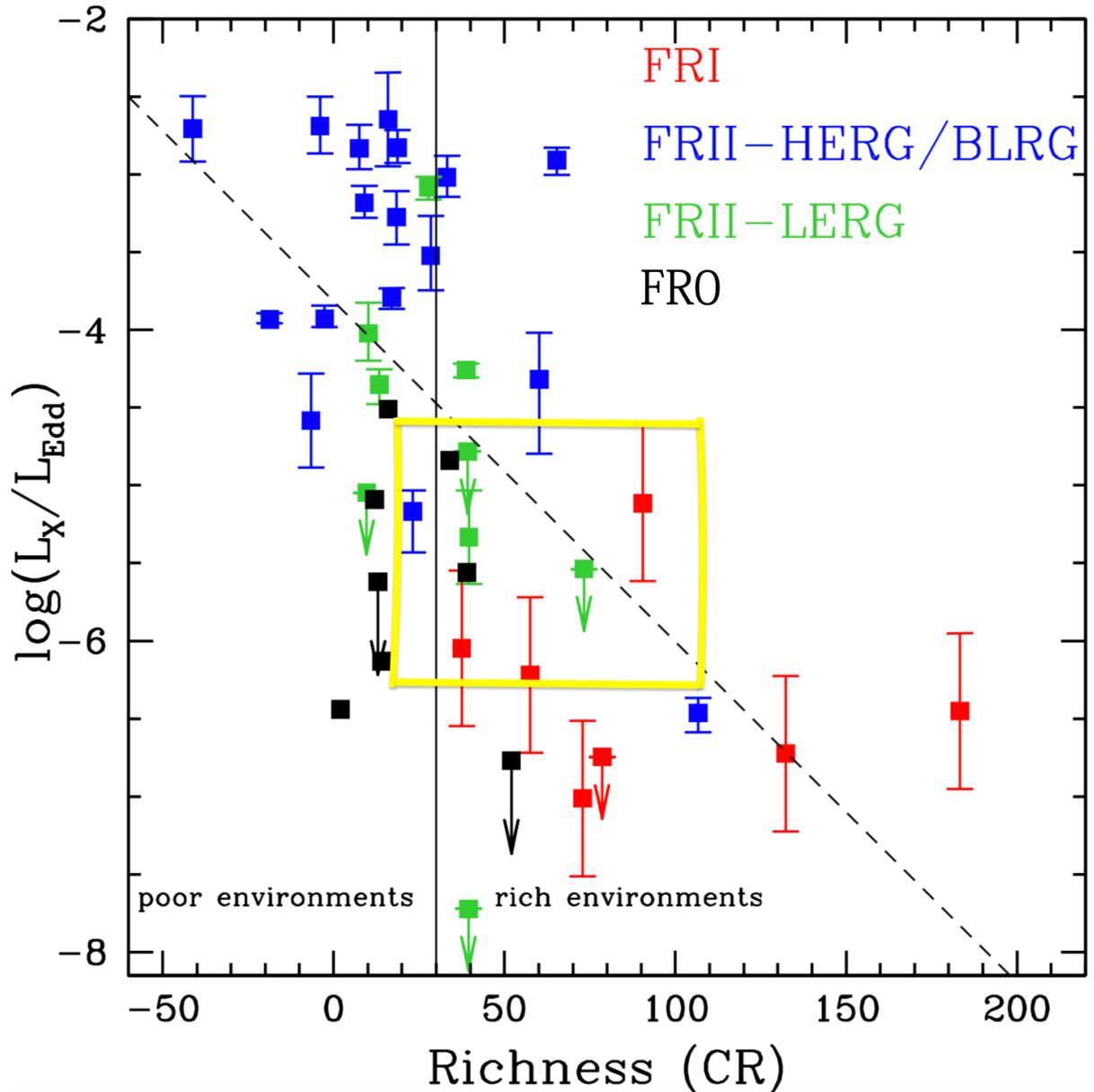

Figure 6: FR0s of Massaro et al (2019) included on the plot of Figure 1. Courtesy Duccio Macconi.

A model prediction is that on average the redshift difference between FR0s and FRIs must be smaller than that between FRIIs and FRIs. Capetti et al. (2020) find no significant redshift difference between FR0s and FRIs but as the numbers increase, the model predicts that average redshift differences between FR0s and FRIs will be less than that between FRIIs regardless of excitation class, and FRIs. Because FRIs evolve from FR0s, as determined by the black hole spin value (i.e. FR0 becomes FRI for *a* > 0.2), the masses of FRIs on average are larger. But the constraints on mass differences are actually tighter. In fact, since FRI LERGs accrete at the lowest rates, they generally do not have enough time to spin their black holes up to high spin (Garofalo 2020). The model predicts that FRIs will on average have black hole masses that are no larger than about 1.5 times the FR0 mass average. This is a powerful prediction that can be tested. Capetti, Massaro & Baldi (2020), compared FR0 masses with masses of samples of FRIs and found the black hole masses on average to be 1.6 times greater for FRIs compared to FR0s.

*-FRI LERGs*

FRI LERGs appear in Figure 1 to distribute themselves at the extremes of the X-ray luminosity/cluster richness relation, having lowest luminosity and highest richness factor. Given the great amount of fuel in rich clusters, Figure 4 allows us to appreciate why such objects live the longest on average in denser environments while that timescale in more isolated environments (if FRI LERGs form there) is less. The radiative efficiency of FRI LERGs is also the lowest among any radio galaxy subgroup as a result of the fact that they are the end of the line in the evolution of powerful FRII HERGs, coming into their own up to billions of years after the triggering of their FRII quasar ancestors. This end of the line picture for FRI LERGs is also compatible with this subclass of the radio galaxy population having the most massive black holes.

3. Conclusions

A powerful constraint on the nature of radio galaxies has emerged in the form of the X-ray luminosity versus cluster richness relation. Despite the low numbers, we have shown that subgroups of the jetted active galaxy phenomenon distribute themselves along this relation in ways that are incompatible with basic assumptions about the triggering of black hole feedback. We have shown how to understand the nature of this distribution from our model perspective for FRII HERGs, FRII LERGs, FR0 LERGs, and FRI LERGs based on a picture in which they are increasingly later stages in the time evolution of radio galaxies. While FRII HERGs are initial states, FRI LERGs are final states, which has both obvious as well as subtle predicted consequences for the redshift, black hole masses, environments, and excitation class of these objects. To our knowledge, no other model can explain in such a straightforward way the distribution of all radio galaxies on the X-ray luminosity/cluster richness plane.

4. Acknowledgments

DG thanks Duccio Macconi for detailed discussion and in particular associated with FRII LERGs and the statistical analysis of their black hole masses and for placing FR0s on Figure 1.